\documentclass[useAMS,usenatbib]{mn2e}

\usepackage{graphicx,times}

\def\solm{M$_{\odot}$}
\def\kms{km s$^{-1}$}
\def\kmskpc{km s$^{-1}$ kpc$^{-1}$}

\def\eg{{\it e.g.\ }}
\def\ie{{\it i.e.\ }}

\title[Regular motions in double bars. I. Double-frequency orbits and loops]
{Regular motions in double bars. \\ I. Double-frequency orbits and loops}

\author[Witold Maciejewski and E. Athanassoula]
{Witold Maciejewski$^{1,2}$\thanks{E-mail:witold@astro.ox.ac.uk} 
and E. Athanassoula$^{2}$\\
$^{1}$Astrophysics Research Institute, Liverpool John Moores University, Twelve Quays House, Egerton Wharf, Birkenhead, CH41 1LD\\
$^{2}$LAM, Observatoire Astronomique de Marseille Provence, CNRS, 2 Place Le Verrier, F-13248 Marseille Cedex 4, France}

\begin{document}

\maketitle

\begin{abstract}
Bars in galaxies are mainly supported by particles trapped around stable
periodic orbits. These orbits represent oscillatory
motion with only one frequency, which is the bar driving frequency, and miss 
free oscillations. We show that a similar situation takes place in double bars:
particles get trapped around parent orbits, which in this case represent 
oscillatory motion with two frequencies of driving by the two bars, and which
also lack free oscillations. Thus the parent orbits, which constitute the
backbone of an oscillating potential of two independently rotating bars, are 
the double-frequency orbits. These orbits do not close in any reference frame, 
but they map onto loops, first introduced by Maciejewski \& Sparke (1997). 
Trajectories trapped around the parent double-frequency orbit 
map onto a set of points confined within a ring surrounding the loop.
\end{abstract}


\section{Introduction}
Double bars in galaxies, also called nested bars or bars within bars, are 
systems with a nuclear bar embedded within a large-scale outer bar. Such
systems appear to be common in galaxies: recent surveys show that 
up to 30\% of early-type barred galaxies contain them (Erwin \& Sparke 2002),
and they are also present in galaxies of types as late as Sbc and Sc (Laine 
et al. 2002). At least 50 double-barred systems are now known (Erwin 2004). 
The relative orientation of the two bars is random (Buta \& Crocker 1993; 
Friedli \& Martinet 1993; Wozniak et al. 1995; Erwin \& Sparke 2002; Laine et 
al. 2002), which is expected if the bars were to rotate with different
pattern speeds. Inner bars, like large bars, are made of relatively old
stellar populations, since they remain distinct in near infrared (Friedli 
et al. 1996). 

Orbits in galaxies with two independently rotating bars do not conserve
the Jacobi integral, and it is a complex dynamical task to explain how such 
systems are sustained. This task has been approached with $N$-body simulations 
(e.g. Friedli \& Martinet 1993; Rautiainen, Salo \& Laurikainen 2002; 
Debattista \& Shen 2007; Heller, Shlosman \& Athanassoula 2007), and through 
studies of trajectories with a selected set of initial conditions (El Zant \& 
Shlosman 2002). Nested bars influence each other throughout their extent:
they can change shapes and accelerate as they rotate
through each other (Maciejewski \& Sparke 2000), and the orbital structure of 
the outer bar can be significantly modified by the inner bar, especially near 
its corotation. In general, the piling up of resonances created by each bar 
may lead to considerable chaotic zones, and, if the fraction of chaotic 
orbits in double bars were large, such systems should not last for long time 
periods. Long-lasting double bars should therefore have a substantial fraction 
of regular orbits, and one should be able to describe their dynamics by means 
similar to those invoked in the case of single bars, \ie in terms of the 
regular orbits that they admit. Such an approach was originally proposed by 
Maciejewski \& Sparke (1997, 2000). In this paper we propose that the backbone 
of the oscillating potential of double bars is made of double-frequency 
orbits, and we explicitly show that these orbits map onto the loops 
invented by Maciejewski \& Sparke. 

In Section 2, we outline in more detail the approach proposed in this paper.
In Section 3, we use the epicyclic approximation to show that two frequencies
on an orbit in the plane of an oscillating potential of a doubly barred galaxy
are indispensable. We also show there that double-frequency orbits map onto
loops in the epicyclic approximation. In Section 4, we analyze particle 
trajectories in double bars in the general non-linear case. We develop a
method of analysis in Fourier space that allows us to distinguish regular
trajectories from chaotic ones, and to recover the three fundamental 
frequencies of the regular orbits. We find that in the general, nonlinear case
double-frequency orbits map onto loops as well, and that they can trap around 
themselves regular orbits.

\section{Conceptual outline}
Particle motion in a potential of double bars belongs to the general problem 
of motion in an oscillating potential (Louis \& Gerhard 1989; Sridhar 1989). In
this particular case, the potential oscillates with the relative period of the 
bars. The best known example of the general problem of motion in an oscillating
potential is the restricted elliptical 3-body problem. Families of closed 
periodic orbits have been found in this last problem, where the test particle 
moves in the potential of a binary star with components on elliptical orbits 
(\eg Broucke 1969). Such families, however, are parametrized by values that 
also characterize the potential (i.e. ellipticity of the stellar orbit and the 
mass ratio of the stars), and the orbital periods there are commensurate with
the oscillation period of the potential. For a given potential, 
these families are reduced to single orbits separated in phase-space. 
The solution for double bars is formally identical, and there an orbit can 
close only when the orbital period is commensurate with the relative 
period of the bars. Howard (1994) found such periodic orbits in double 
bars with periods equal to a multiplicity of the bars' relative period.
Such orbits are separated in phase-space and therefore
families of closed periodic orbits are not sufficient to provide orbital 
support for nested bars. 

Binney \& Spergel (1982, 1984) developed a formalism that classifies the 
orbits based on the frequencies at which they oscillate. The Fourier transform 
of any dynamical variable on a regular orbit (position, velocity) should 
return a line spectrum. Regular orbits in the plane of a stationary galaxy 
correspond to oscillations with two frequencies: the angular frequency, 
and the epicyclic frequency -- the orbit's fundamental frequency. Closed 
periodic orbits are a particular case of regular orbits. They oscillate with 
one frequency only, whose inverse is the period of the orbit. In general, a 
single frequency in the system may appear when the two frequencies above are
commensurate, or when oscillations with one of the frequencies are missing. 
A rotating asymmetry (\eg a bar) in the system constitutes a periodic 
driving force. As in the case of a driven harmonic oscillator, the orbits 
always respond to this driving force by oscillations with the frequency of 
the driver. These oscillations are always present, and if the orbit is to
close in the frame rotating with the bar, they must be the only oscillations, 
i.e. the free oscillations with the epicyclic frequency must be absent.

The oscillating potential of double bars changes 
periodically with the relative period of the bars.
Since a Hamiltonian system in $N$ dimensions with a periodically 
varying potential is equivalent to an autonomous system in $N+1$ dimensions 
(see \eg Lichtenberg \& Lieberman 1992, Louis \& Gerhard 1988), regular 
motions in the plane of a doubly barred galaxy should correspond to 
oscillations with three frequencies. As before, Fourier analysis of a closed 
periodic orbit will return only one frequency.
Aside from oscillations with three frequencies,
and with one frequency only, there remain oscillations with two frequencies. 
Since two bars in a doubly barred galaxy constitute two periodic driving
forces, oscillations with two frequencies should characterize response of
the system to this periodic driving, when free oscillations are absent. Thus,
the role of double-frequency orbits in double bars should be equivalent to the 
role of closed periodic orbits in a single bar. In the unchanging potential of 
a single bar, regular orbits are trapped around stable closed periodic orbits,
which constitute the backbone of that potential. Here we postulate that stable 
double-frequency orbits constitute the backbone of an oscillating potential of
double bars, and trap around themselves regular orbits with the maximum number
of frequencies, which, in this case, is three.

The approach outlined above indicates that oscillating potentials can be 
supported by orbits that do not have to be closed and periodic. With this 
approach we overcome the problem of scarcity of closed periodic 
orbits that hampered previous searches for support of oscillating potentials.
The concept of closed periodic orbit is too limiting for the investigation 
of oscillating systems, and we replace it by another, broader description of 
particle motion, which still captures its regularities. Our description, 
based on double-frequency orbits which are not closed and periodic, can
provide the backbone of an oscillating system for the same reason for which 
closed periodic orbits in stationary potentials do, and account for longevity 
of double bars.

Double-frequency orbits are difficult to visualize, because they do not close 
in any reference frame. However, in this paper we will show that they can be
mapped onto closed curves, which are much easier to visualize. The mapping is 
done
by recording the position of a particle on a double-frequency orbit at every
consecutive alignment of the bars. Thus particle's position is written only
when the oscillating galaxy periodically regains one specified shape. In this
paper we show that positions of particles on a double-frequency orbit recorded
at every consecutive alignment of the bars populate a closed curve. By 
construction, this curve transforms into itself at each consecutive 
alignment of the bars. Thus a set of particles populating such a curve
oscillates periodically with the oscillating potential. Maciejewski \& Sparke 
(1997) discovered such curves in a potential with a double bar, and they called
them loops. Loops can be used as a convenient descriptor of orbital structure 
in an oscillating potential. Like closed periodic orbits in stationary 
potentials, they can indicate the zones that can be supported by regular 
motions. In the remaining sections we will show that double-frequency orbits 
can form the backbone of oscillating potentials, and that loops make a unique 
tool for studying such orbits.

\section{The epicyclic solution }

\subsection{The general epicyclic solution for any number of bars}

If a galaxy has a bar that rotates with a constant pattern speed,
it is convenient to study particle orbits in the reference frame rotating
with the bar. If two or more bars are present and each rotates with 
its own pattern speed, there is no reference frame in which the potential
remains unchanged. In order to point out formal similarities in solutions 
for one and many bars, we solve the linearized equations of motion in the 
inertial frame. As we show below, this is equivalent to the solution in any 
rotating frame 
and the transformation is particularly simple: in the rotating frame the 
centrifugal and Coriolis terms are equivalent to the Doppler shift of the
angular velocity. 

In a frame rotating with a constant pattern speed $\Omega_B$, the equation 
of motion of a particle in a potential $\Phi$ can be written as 
\begin{displaymath}
\ddot{\bf r'} = -\nabla \Phi  - 2 ({\bf \Omega_B} \times \dot{\bf r'})
- {\bf \Omega_B} \times ( {\bf \Omega_B} \times {\bf r'}),
\end{displaymath}
where the position of the particle in the rotating frame is indexed by prime.
In polar coordinates in the plane of the galactic disc, its $R'$ and 
$\varphi'$ components are
\begin{eqnarray}
\ddot{R'} - R' ( \dot{\varphi'} + \Omega_B )^2
& = & -\frac{\partial \Phi}{\partial R}, \\
R' \ddot{\varphi'} + 2 \dot{R'} ( \dot{\varphi'} + \Omega_B )
& = & -\frac{1}{R} \frac{\partial \Phi}{\partial \varphi}.
\end{eqnarray}
Since the angular velocity $\dot{\varphi'}$ in the rotating frame is related 
to the angular velocity $\dot{\varphi}$ in the inertial frame by 
\begin{displaymath}
\dot{\varphi}=\dot{\varphi'}+\Omega_B,
\end{displaymath}
and radial coordinates in both frames are equal ($R=R'$), equations (1) and 
(2) in the inertial frame take the form 
\begin{eqnarray}
\label{eqmri}
\ddot{R} - R \dot{\varphi}^2
& = & -\frac{\partial \Phi}{\partial R}, \\
\label{eqmfi}
R \ddot{\varphi} + 2 \dot{R} \dot{\varphi}
& = & -\frac{1}{R} \frac{\partial \Phi}{\partial \varphi} ,
\end{eqnarray}
which are the $R$ and $\varphi$ components of the equation of motion in the 
inertial frame: 
\begin{displaymath}
 \ddot{\bf r} = -\nabla \Phi .
\end{displaymath}
This equivalence remains true when vertical motion is included in three 
dimensions. For the rest of this section we adopt the inertial frame, 
in which the $R$ and $\varphi$ components of the equation of motion take 
a particularly simple form of (\ref{eqmri}, \ref{eqmfi}). 

We want to linearize equations (\ref{eqmri}) and (\ref{eqmfi}) for departures
from circular motion in a nearly axisymmetric potential. We thus perform two 
expansions simultaneously. One is a linearization of the quantities describing 
the motion by their expansion to first-order terms in $R-R_0$, where $R_0$ 
is the radius of the circular orbit in the axisymmetric
potential. The other is expansion of the potential in a Fourier series, with 
the zeroth-order term being the axisymmetric component. In both expansions, 
zeroth-order terms are indexed with $0$ and
first-order terms are indexed with $I$. These expansions can be written as
\begin{eqnarray}
\label{e5}
R(t)              & = & R_0 + R_I(t) , \\
\label{e6}
\varphi(t)        & = & \varphi_{00} + \Omega_0 t + \varphi_I(t) , \\
\label{e7}
\Phi(R,\varphi,t) & = & \Phi_0(R) + \Phi_I(R,\varphi,t) ,
\end{eqnarray}
where second- and higher-order terms are neglected. In the inertial frame,
the angle coordinate $\varphi$ of a particle in the zeroth approximation is 
$\varphi_0 = \varphi_{00} + \Omega_0 t$, where $\varphi_{00}$ is the value 
of $\varphi_0$ at $t=0$, and where the angular velocity $\Omega_0$ on the circular 
orbit of radius $R_0$ relates to the  potential $\Phi_0$ through the zeroth 
order of (\ref{eqmri}):
$\Omega_0^2 = (1/R_0) (\partial \Phi_0 / \partial R) \; |_{R_0}$.
The zeroth order of (\ref{eqmfi}) is identically equal to zero. In the zeroth
order approximation, the axisymmetric part of the potential, $\Phi_0$, is 
time-independent. The first order term $\Phi_I$, describing asymmetry in the 
potential, may depend on time. 

In the expansion defined above, the first order parts of (\ref{eqmri}) 
and (\ref{eqmfi}) take respectively the forms
\begin{eqnarray}
\label{rlin}
\ddot{R_I}  -  4 A \Omega_0 R_I  -  2 R_0 \Omega_0 \dot{\varphi_I}  & = &  
-\frac{\partial \Phi_I}{\partial R} \; |_{R_0,\varphi_0}, \\
\label{flin}
R_0 \ddot{\varphi_I}  + 2 \Omega_0 \dot{R_I}  & = &  
-\frac{1}{R_0} \frac{\partial \Phi_I}{\partial \varphi}  \; |_{R_0,\varphi_0},
\end{eqnarray}
where $A$ is the Oort constant defined by 
$ 4 A \Omega_0 = \Omega_0^2 - \frac{\partial^2 \Phi_0}{\partial R^2} |_{R_0}$.

Furthermore, we assume that, to the first order, the departure
of the barred potential from axial symmetry can be described by a term
$\cos (2 \varphi)$. If $N$ bars, indexed by $i$, rotate independently 
as solid bodies with angular velocities $\Omega_i$, the 
time-dependent first-order correction $\Phi_I$ to the potential can be
written as
\begin{equation}
\label{phi1}
\Phi_I(R,\varphi,t) = \sum_i^N \Psi_i(R) \cos[ 2 (\varphi - \Omega_i t) ] ,
\end{equation}
where the radial dependence $\Psi_i(R)$ has been separated from the angular 
dependence. The lack of phase in the above trigonometric functions means that 
we define as $t=0$ the time when all the bars are aligned. Derivatives of 
(\ref{phi1}) enter the right-hand sides of (\ref{rlin}) and (\ref{flin}), 
which after introducing 
\begin{equation}
\omega_i = 2 (\Omega_0 - \Omega_i) ,
\label{omom}
\end{equation}
take the form
\begin{equation}
\label{rlin1}
- \ddot{R_I}  +  4 A \Omega_0 R_I  +  2 R_0 \Omega_0 \dot{\varphi_I} =  
\sum_i^N \frac{\partial \Psi_i}{\partial R} \; |_{R_0} \cos(\omega_i t + 2 \varphi_{00})
\end{equation}
\begin{equation}
\label{flin1}
R_0 \ddot{\varphi_I}  + 2 \Omega_0 \dot{R_I} =
\frac{2}{R_0} \sum_i^N \Psi_i(R_0) \sin (\omega_i t + 2 \varphi_{00}).
\end{equation}

In order to solve the set of equations (\ref{rlin1},\ref{flin1}), one can 
integrate (\ref{flin1}) and get an expression for $R_0 \dot{\varphi_I}$, 
which furthermore can be substituted into (\ref{rlin1}). This substitution
eliminates $\varphi_I$, and one gets a single second order equation for $R_I$,
which can be written schematically as 
\begin{equation}
\label{ddotR}
\ddot{R_I} + \kappa_0^2 R_I = \sum_i^N M_i \cos (\omega_i t + 2 \varphi_{00})
                              + C_{\varphi},
\end{equation}
where 
$ M_i = - \frac{4 \Omega_0 \Psi_i}{\omega_i R_0}
        - \frac{\partial \Psi_i}{\partial R}_{| R_0} $,
$\kappa_0^2 = 4 \Omega_0 ( \Omega_0 - A )$, and $C_{\varphi}/2\Omega_0$ is
the integration constant that appears after integrating (\ref{flin1}). Below,
we will show that $C_{\varphi}$ can be removed by redefining the guiding
radius. What is left of (\ref{ddotR}),
is the equation of a harmonic oscillator with multiple driving terms. Thus
in the linear approximation, the orbital motion of a particle in the potential
of rotating bars is equivalent to the motion of a driven harmonic oscillator.
The general solution of (\ref{ddotR}) is well known and can be written as
\begin{equation}
\label{r1}
R_I (t) = C_1 \cos (\kappa_0 t + \delta) 
   + \sum_i^N P_i \cos(\omega_i t + 2 \varphi_{00})
   + C_{\varphi}/\kappa_0^2 ,
\end{equation}
where the coefficients $P_i$ are functions of $M_i$. 
The first term in the right hand side of (\ref{r1}) is the
solution of the homogeneous part of  
(\ref{ddotR}), and it corresponds to a free oscillation
with the local epicyclic frequency $\kappa_0$ around the circular orbit in the
axisymmetric potential. The coefficient $C_1$ is unconstrained. The terms in 
the sum over $i$ in (\ref{r1}) are the particular solutions of the 
inhomogeneous 
equation (\ref{ddotR}), and they describe oscillations driven by the bars. 
Note that the frequency of the driving force (shifted to the frame of the
rotating guiding centre) is always present in the oscillations. Thus, orbits
in potentials with $N$ independently rotating bars will oscillate with at
least $N$ frequencies, unless the shifted frequencies are commensurate. 

The formula for  $\varphi_I(t)$ can be obtained by substituting (\ref{r1}) 
into the time-integrated (\ref{flin1}). As a result, one gets
\begin{equation}
\label{phi1dot}
\dot{\varphi_I} =  C_2 \cos (\kappa_0 t + \delta)
                 + \sum_i^N Q_i \cos(\omega_i t + 2 \varphi_{00})
                 - \frac{2 A C_{\varphi}}{\kappa_0^2 R_0},
\end{equation}
where again $C_2$ is unconstrained and the $Q_i$ are determined by the 
coefficients of the equations above. Note that the integration constant 
$C_{\varphi}/\kappa_0^2$ in (\ref{r1}) can be incorporated into $R_0$ in 
(\ref{e5}), giving the new guiding radius $R_0 + C_{\varphi}/\kappa_0^2$. 
Since $A = -\frac{1}{2} R_0 \frac{d\Omega_0}{dR} | R_0$, expanding $\Omega$ 
to the first order gives the angular velocity at this new radius
\begin{displaymath}
\Omega_0 |_{R_0 + C_{\varphi}/\kappa_0^2} \; = \;
\Omega_0 |_{R_0} + \frac{C_{\varphi}}{\kappa_0^2} \frac{d\Omega_0}{dR}|_{R_0} 
\; = \;
\Omega_0 |_{R_0} - \frac{2 A C_{\varphi}}{\kappa_0^2 R_0} .
\end{displaymath}
Thus the integration constant $ - \frac{2 A C_{\varphi}}{\kappa_0^2 R_0}$ in  
(\ref{phi1dot}) represents the correction to the zeroth order angular velocity 
$\Omega_0$ incurred by a change of the guiding radius $R_0$ by 
$C_{\varphi}/\kappa_0^2$, which is the integration constant in (\ref{r1}).
Hence both integration constants can be incorporated by modifying the guiding
radius in the epicyclic approximation. Eventually, in the epicyclic 
approximation, the most general solution for particle motion in the symmetry
plane of a potential with $N$ independently rotating bars is
\begin{equation}
\label{r1f}
R_I (t) = C_1 \cos (\kappa_0 t + \delta) + \sum_i^N  P_i \cos(\omega_i t + 2 \varphi_{00}) ,
\end{equation}
\begin{equation}
\label{f1f}
\varphi_I(t) = \frac{C_2}{\kappa_0} \sin (\kappa_0 t + \delta) + \sum_i^N \frac{Q_i}{\omega_i} \sin(\omega_i t + 2 \varphi_{00}) + \rm const ,
\end{equation}
where $R_I$ and $\varphi_I$ are polar coordinates in the inertial frame. The 
integration constant in (\ref{f1f}) is an unconstrained parameter. From
(\ref{r1f}) and (\ref{f1f}) one can see that up to 
$N+1$ independent frequencies are present in this motion.
 
\subsection{Closed periodic orbits in a single bar}
In a potential with a single bar there is only one term in the sums over $i$
in (\ref{r1f}) and (\ref{f1f}). Therefore in general two frequencies are 
present in orbital motion, as expected for motion in a symmetry plane of a
static potential. However, closed periodic orbits have only one frequency. 
Oscillations with the frequency of the rotating bar (corrected for the 
angular frequency of the guiding centre) are always present in orbital motion, 
analogously to oscillations with the frequency of the driving force in a driven
harmonic oscillator. The linear approximation breaks down when the
frequency of free oscillations becomes commensurate 
with the frequency of the driving force. 
Therefore, in this framework, in order for the orbits 
to close, the amplitude of free oscillations with frequency $\kappa_0$ has to 
be zero. For $C_1=C_2=0$ and $N=1$, equations (\ref{r1f}) and 
(\ref{f1f}) describe closed periodic orbits in the linearized problem of a 
particle motion in a single bar.

Consider, for the case where free oscillations are absent, the change
in the values of $R_I$ and $\varphi_I$ after the particle's 
guiding centre returns to its starting point in the frame corotating 
with the bar. This happens after time period $2\pi / (\Omega_0 - \Omega_1)$.
Replacing $t$ by $t + 2\pi / (\Omega_0 - \Omega_1)$ leads to
\begin{eqnarray*}
R_I & = &
P_1 \cos[ \omega_1 (t + \frac{2\pi}{\Omega_0 - \Omega_1}) + 2 \varphi_{00}] \\
& = & 
P_1 \cos( \omega_1 t + 4 \pi + 2 \varphi_{00} ).
\end{eqnarray*}
Thus, after time $2\pi / (\Omega_0 - \Omega_1)$ the solution for $R_I$
returns to its starting value. The same holds true for $\varphi_I$ and the 
orbit closes in the bar frame. Thus, $2\pi / (\Omega_0 - \Omega_1)$ is the 
full period of this particle.

\subsection{Closed periodic orbits in double bars}
While the potential of a single bar is stationary in the bar's frame,
that of a galaxy
with two independently rotating bars is not stationary in any reference frame.
For two bars there are two terms in the sums over $i$ in (\ref{r1f}) and 
(\ref{f1f}), and therefore even if the amplitude of free oscillations with 
the frequency $\kappa_0$ is zero, two frequencies are present in the orbital
motion. They are the frequencies of the driving forces imposed by the two tumbling
bars, and oscillations with these two frequencies have always non-zero 
amplitude. Orbits with two frequencies do not close, therefore closed periodic 
orbits should not be the fundamental orbits in double bars.

However, closed periodic orbits do exist in double bars, and in the linear
approximation this happens when the frequencies in the sums over $i$ in 
(\ref{r1f}) and (\ref{f1f}) are commensurate, \ie when $n\omega_1=m\omega_2$ 
for $n,m$ integer. In that case there is effectively only one independent 
frequency and the orbit closes. However, this happens only for a discrete 
set of values of the angular velocity of the guiding centre 
\begin{equation}
\label{cl2b}
\Omega = \frac{m\Omega_1-n\Omega_2}{m-n} .
\end{equation}
Therefore in the epicyclic approximation there is an infinity of closed periodic 
orbits in double bars, but these orbits do not form families continuous with 
the radius of the guiding centre. Although formally any orbit in double bars, 
which responds to two driving forces by oscillating with two
frequencies, can be approximated by a closed periodic orbit of angular
frequency (\ref{cl2b}), such an orbit closes after very many bar periods,
and describing motions in double bars in terms of such closed periodic
orbits is not very convenient. Moreover, although any orbit oscillating with 
two frequencies has a closed periodic orbit in its infinitesimal vicinity with 
sufficiently large $m$ and $n$, such a closed periodic orbit cannot be 
considered to be a parent orbit, because orbits with two independent frequencies are not 
constituted by free oscillations around that orbit. Finally, in the non-linear 
regime, Howard (1994) found that closed periodic orbits in doubly barred 
potentials, with periods predicted by (\ref{cl2b}) are often unstable, and 
therefore they cannot serve as parent orbits. Thus, closed periodic orbits are 
unlikely to provide sufficient backbone supporting double bars. Below we 
present a more convenient way serving this purpose.

\subsection{Double-frequency orbits in double bars. Loops}
In Sect.3.2 we showed that in the case of a single bar, when there are no
free oscillations, the linear solution (\ref{r1f}) and (\ref{f1f}) returns 
only one oscillation frequency, which indicates a closed periodic orbit.
In a solution for double bars, two independent frequencies are present in 
general, hence one should expect that {\it double-frequency} orbits take place 
of closed periodic orbits there. Such orbits do not close 
in any reference frame, because when a term from one bar in (\ref{r1f}) and 
(\ref{f1f}) returns to its initial value, the term from the other bar does 
not. 

However, consider the change in value of $R_I$ and
$\varphi_I$ after time $\pi / (\Omega_2 - \Omega_1)$, which is the relative
period of the bars. Every such time interval, the mass distribution in the 
galaxy and its potential are the same in the frame rotating with either bar. 
With amplitudes of free oscillations in (\ref{r1f}) set to zero one gets
\begin{eqnarray*}
R_I & = &
P_1 \cos[ \omega_1 (t + \frac{\pi}{\Omega_2 - \Omega_1}) + 2 \varphi_{00}] \\
& + &
P_2 \cos[ \omega_2 (t + \frac{\pi}{\Omega_2 - \Omega_1}) + 2 \varphi_{00}] \\
& = &
P_1 \cos( \omega_1 t + 2 \pi \frac{\Omega_0 - \Omega_1}{\Omega_2 - \Omega_1} + 2 \varphi_{00} ) \\
& + &
P_2 \cos( \omega_2 t + 2 \pi \frac{\Omega_0 - \Omega_2}{\Omega_2 - \Omega_1} + 2 \varphi_{00} ) \\
& = &
P_1 \cos( \omega_1 t + 2 \pi + 2 \varphi_{01} ) +
P_2 \cos( \omega_2 t + 2 \varphi_{01}),
\end{eqnarray*}
where $\varphi_{01} = \varphi_{00} + \pi \frac{\Omega_0 - \Omega_2}{\Omega_2 - \Omega_1}$. The same result can be obtained for $\varphi_I$. This means that
the time transformation $t \rightarrow t + \pi / (\Omega_2 - \Omega_1)$ is
equivalent to the change in the starting position angle of a particle from
$\varphi_{00}$ to $\varphi_{01}$. Consider the motion of a set of particles
that have the same guiding radius $R_0$, but start at various position
angles $\varphi_{00}$. This is a one-parameter set, therefore in the disc
plane it is represented by a curve, and because of continuity of (\ref{r1f})
and (\ref{f1f}) and their periodicity at a fixed $t$, this curve is closed.
After time
$\pi / (\Omega_2 - \Omega_1)$, a particle starting at angle $\varphi_{00}$ will
take the place of the particle which started at $\varphi_{01}$, a particle
starting at $\varphi_{01}$ will take the place of another particle from this
curve and so on. The whole curve will regain its shape and position every
$\pi / (\Omega_2 - \Omega_1)$ time interval, although positions of particles
on the curve will shift. This curve is the epicyclic approximation to the
{\it loop} introduced by Maciejewski \& Sparke (1997, 2000): a curve made of 
particles moving in a given potential, such that the curve returns
periodically to its original shape and position. In the case of two bars, 
the period is the relative period of the bars, and in the epicyclic 
approximation the loop is made out of particles having the same guiding 
radius $R_0$. Particles on the loop respond to the periodic driving force 
from the two bars, but they lack any free oscillations. Note that the 
appearance of the loops is the same, no matter which bar's reference frame
is used. Thus, loops are a convenient descriptor of orbital structure in an 
oscillating potential. An example of a set of loops in a doubly barred galaxy 
in the epicyclic approximation was given by Maciejewski \& Sparke (1997).
The loops occupy there a significant part of the disc.

Note that the linear solution presented here gives a prescription for 
constructing a loop in a general case as well. If a particle is on a 
double-frequency 
orbit, then the loop onto which the orbit maps can be constructed by
registering the position of the particle every relative period of the bars,
until  the whole $2\pi$ range of angular coordinates is densely populated.
However, this construction requires following the particle for many relative
periods of the bars, hence it will likely recover loops that are maps of 
stable double-frequency orbits only.

\begin{figure}
\centering
\vspace{-3mm}
\includegraphics[width=1.05\linewidth]{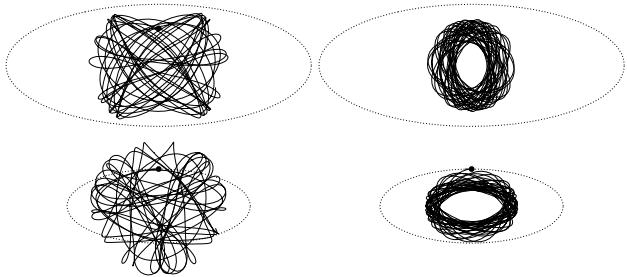}
\vspace{-85mm}
\caption[]{Two example trajectories (one in the two left panels, one in 
the two right ones) of a particle that moves in the potential of two 
independently rotating bars from Model 1 of Maciejewski \& Sparke (2000). 
Large dots mark the starting position of the particle at 1.7-kpc distance 
from the centre of the galaxy, on 
the minor axis of the bars, at the moment when the corresponding axes of 
both bars overlap. The initial particle's velocity vector is in both examples
perpendicular to the minor axis, and its value equal to $v_x=220$ \kms\ (left 
panels) and $v_x=150$ \kms\ (right panels). The particle is followed for 10 
relative 
periods of the bars, and its trajectory is displayed in the frame corotating 
with the big, outer bar (top panels); and with the small, inner bar (bottom 
panels). Each bar is outlined in its own reference frame by a dotted line.}
\label{f1}
\end{figure}

\section{Full nonlinear solution for particle motion in nested bars}
\subsection{Three types of trajectories}
Tools and concepts useful in the search for ordered motions in double bars are 
best introduced through the inspection of particle trajectories in such 
systems. We use the potential of Model 1 defined in Maciejewski \& Sparke 
(2000), so that our results can be directly compared to the results there.
In order to describe gravitational potential in their models, Maciejewski \& 
Sparke use the analytical formulae originally proposed by Athanassoula (1992).
The formulae comprise the modified Hubble profile representing the bulge,
the Kuzmin-Toomre profile representing the disc, and the Ferrers formula
describing each bar. The exponent in the Ferrers formula is 2. In Model 1
of Maciejewski \& Sparke, the outer, main bar is 7 kpc long, and the inner 
bar is 60\% in size of the outer one. The axial ratio of each bar is 2.5. The
quadrupole moment of the outer bar is only $2.25 \times 10^{10}$ \solm kpc$^2$
in Model 1, and the mass of the inner bar is 60\% of that of the outer one,
hence both bars in this model are rather weak. Pattern speeds of the bars 
are not commensurate ($\Omega_1=24.03$ \kmskpc, $\Omega_2=41.90$ \kmskpc),
and indicate rapidly rotating bars: the corotation radius of each bar is
about 20\% larger than its semi-major axis. The Inner Lindblad Resonance 
of the outer bar is located at 2.9 kpc, hence there is no resonant coupling
between the bars in this model.

Consider a particle moving in this potential inside the corotation of the 
inner bar. Simple experiments exploring a limited range of initial conditions
-- with the particle starting on the minor axis of the bars at the moment when
the axes overlap and with velocity vector perpendicular to that axis -- show
that if the initial velocity is small enough, the particle usually remains 
bound. Its trajectory may look regular or irregular. To illustrate this,
in Figures 1 and 2
we present three trajectories of a particle starting at 1.7-kpc distance from 
the centre of the galaxy with three different initial velocities. A typical 
irregular trajectory is shown in the left panels of Fig.1 (starting velocity
$v_x=220$ \kms). Since the trajectory depends on the 
reference frame, it is plotted twice, once for the reference frame of each bar.
However, particle trajectories usually look more regular: an example is given
in the right panels of Fig.1 (starting velocity $v_x=150$ \kms). These 
trajectories look like if they were trapped around some stable regular orbit. 

\begin{figure}
\centering
\vspace{-5mm}
\includegraphics[width=1.05\linewidth]{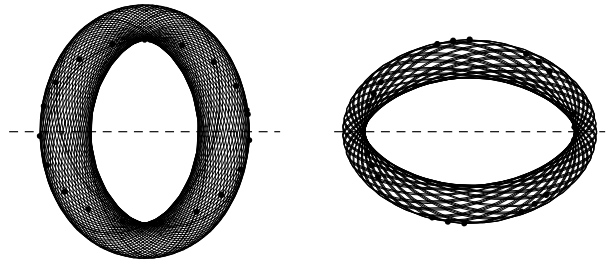}
\vspace{-5cm}
\caption[]{Most regular trajectory that starts at 1.7-kpc distance from the 
centre of the doubly barred potential from Model 1 of Maciejewski \& Sparke 
(2000), followed for 20 relative periods of the bars, and written in the frame 
corotating with the big bar (left), and with the small bar (right). The 
major axis of each bar in its own reference frame is marked by the dashed 
line. Dots mark positions of the particle at every alignment of the bars (see
Sect.4.4).}
\label{f2}
\end{figure}

Fine adjustments of the initial velocity lead to a highly harmonious trajectory
(Fig.2, starting velocity $v_x=180$ \kms), which appears even more regular than
the trajectory from the right panels of Fig.1. Moreover, this is the most 
regular trajectory that we can find by varying the initial velocity. In 
particular, we do not find any closed periodic orbits that this trajectory
can be related to. Below we will formalize the meaning of being `most regular' 
in this context. The trajectory from Fig.2 looks like a loop orbit in a 
potential of a single bar\footnote{in the existing nomenclature, which we
continue to use, a loop orbit is an orbit in a time-independent potential, 
and it has no relation to a loop, which is a map of a double-frequency orbit 
in a time-dependent oscillating potential} (see e.g. fig.3.7a in Binney \& 
Tremaine 1987). In Sect.4.2, we will show that this morphological 
similarity occurs because in both cases orbital oscillations with two 
frequencies are involved. The loop orbit in a single bar forms when a particle 
oscillates around a closed periodic orbit. Therefore, it involves two 
frequencies: the frequency of the free oscillation, and the frequency of 
driving by the bar. In Sect.4.2.1 we will demonstrate that 
in the trajectory from Fig.2, the two frequencies are the
frequencies of driving by the two bars, while the free oscillation is absent.
This is the reason why the trajectory from Fig.2 appears most regular: adding
the free oscillation with its own frequency will make the appearance of the
trajectory less regular.

\begin{figure}
\centering
\vspace{-2mm}
\includegraphics[width=1.05\linewidth]{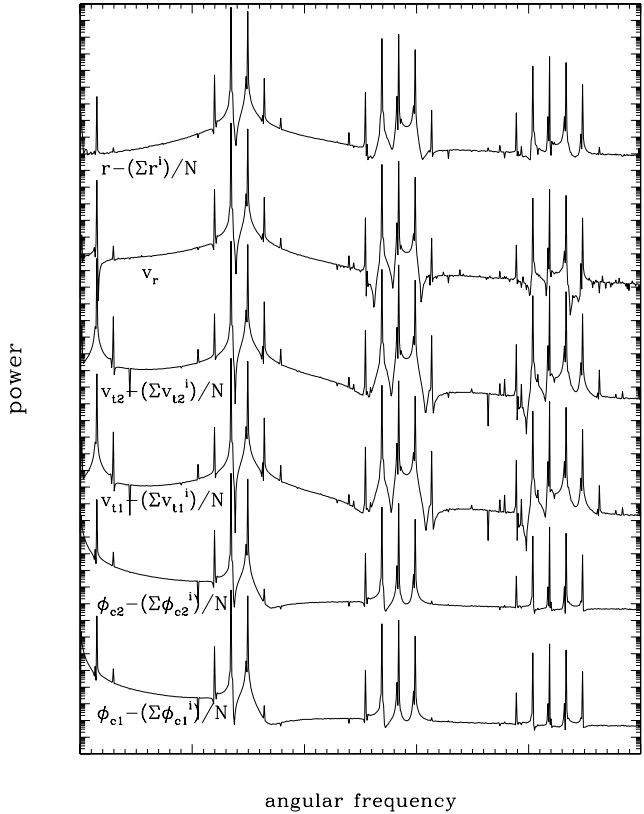}
\vspace{-9mm}
\caption[]{One-sided power spectral density (power spectra) from fast Fourier 
transform of all six dynamical variables described in the text on an exemplary 
regular trajectory. The vertical axis is logarithmic, with the value increasing
by one order of magnitude every large tick mark. For clarity, the spectra are 
shifted in power from their neighbours by 7 large tick marks each.}
\label{f3}
\end{figure}

\subsection{Orbital frequencies}
In order to see what frequencies are present in the motion of a particle on a 
given trajectory, we Fourier-analyse any dynamical variable on this trajectory.
Fast Fourier transform method
has been employed, and the trajectory was traced for 100 alignments of
the bars, with dynamical variables sampled $N=2^{13}$ times at equal time 
intervals. We transformed the polar coordinates of the 
particle, as well as its polar velocity components. The values of the radial 
velocity $v_r$ entered the transformation directly, while the radial 
coordinate $r$ had its average over $N$ samplings subtracted before it was 
transformed. Similarly, the average over $N$ samplings of the tangential 
velocity $v_t$ was subtracted from its values before they entered the Fourier 
transform. In order to transform the tangential coordinate $\varphi$, we also 
had to subtract the component reflecting its monotonic growth. In effect the 
following values were transformed: $v_r$, $r - (\Sigma r^i) / N$, 
$v_t - (\Sigma v_t^i) / N$, 
$\varphi_c - (\Sigma  \varphi_c^i) / N$, where 
$\varphi_c = \varphi - \bar{\Omega}t$, $t$ is the time since the moment when
the particle started, $\bar{\Omega} = (\varphi^N - \varphi^1)/T$, and $T$ is 
the total time for which the particle has been traced. The sum $\Sigma$ is 
over all $N$ samplings, and the upper index counts the samplings. Moreover, 
the angular coordinate $\varphi$ and the tangential velocity $v_t$ were 
transformed for the frame of reference of each bar 
($\varphi_{c1}$, $\varphi_{c2}$ and $v_{t1}$, $v_{t2}$), hence 6 dynamical
variables were transformed in total.

\begin{figure*}
\centering
\vspace{-2mm}
\rotatebox{-90}{\includegraphics[width=0.74\linewidth]{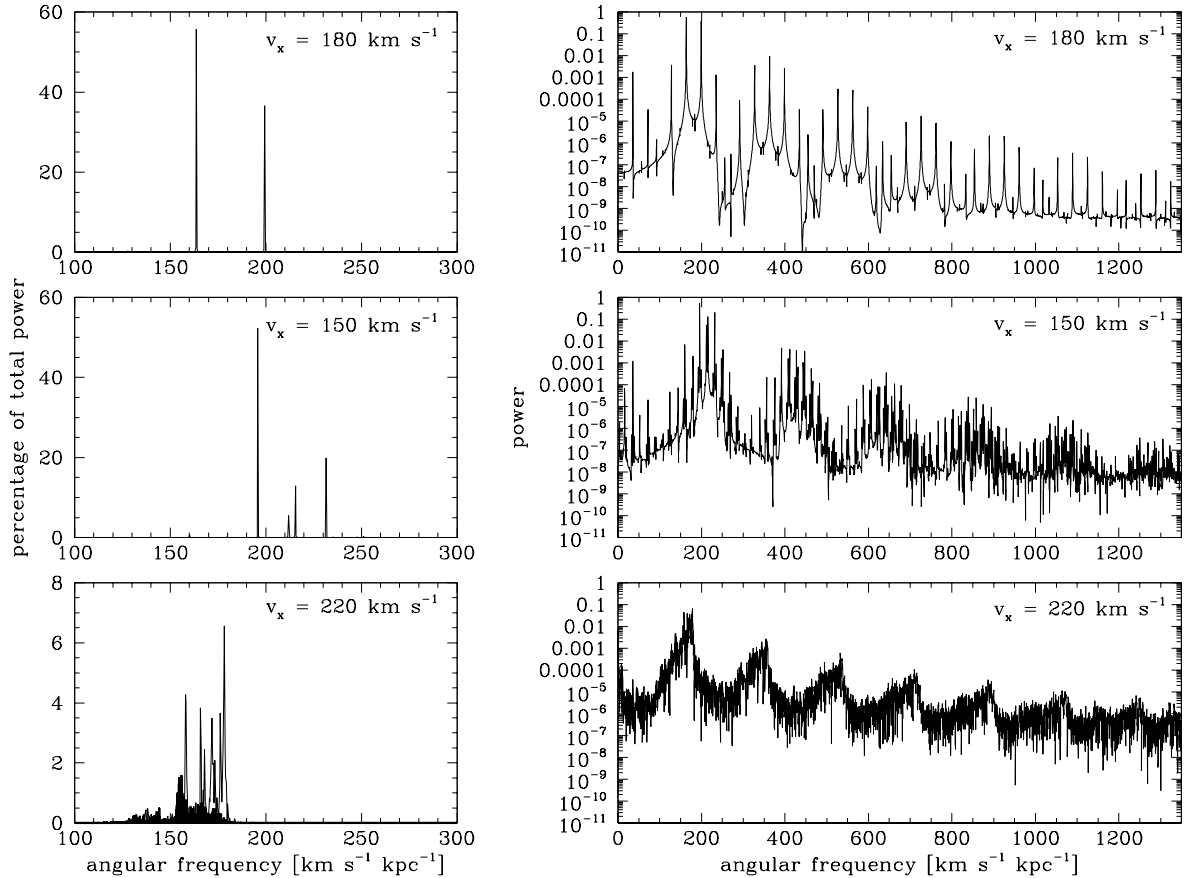}}
\vspace{-9mm}
\caption[]{Power spectra of the radial coordinate on three representative
trajectories, shown in linear (left-hand panels) and logarithmic (right-hand
panels) scale. The plots in the top row show the power spectrum of the 
double-frequency orbit from Fig.2. The plots in the middle and bottom rows
show, respectively, power spectra of the trajectories from the right-hand and 
left-hand panels of Fig.1. The particle was followed for 100 relative periods 
of the bars, with dynamical variables sampled $N=2^{13}$ times at equal time 
intervals. In the bottom-left panel, the thick line marks an additional power 
spectrum, for the same initial conditions, but with the particle followed for 
600 relative periods of the bars, with dynamical variables sampled $N=2^{16}$ 
times.}
\label{f4}
\end{figure*}

In Fig.3, we present one-sided power spectral density (for brevity called here 
power spectra) from fast Fourier transform for each variable on an exemplary 
regular trajectory. They are essentially line spectra, as most of the power 
resides in a few separated peaks. Note the quality of the spectra: the power 
density in the  peaks is several orders of magnitude higher than that in the 
baseline. Power spectra
of all six dynamical variables transformed have the same distribution and 
similar strengths of major peaks. Only spectra of the angular coordinate
$\varphi$ are of slightly lower quality, with higher baseline, and power
increasing towards the lowest frequencies. This is likely the effect of 
inaccuracies in double corrections applied to the measured value of those 
variables. Even in these spectra, however, the peaks are at least two orders 
of magnitude in power over the highest values of the baseline. In the rest
of this paper, we focus on the analysis of the spectrum of the radial 
coordinate only, for which our Fourier algorithm returns data of best 
quality.

\subsubsection{Most regular trajectory as a double-frequency orbit}
In Fig. 4, we present power spectra for the three trajectories from Figures 1 
and 2. We display the power density in both the linear scale (left panels), 
and the logarithmic scale (right panels). The power spectrum for the most 
regular trajectory from Fig.2 is shown in the top row. It shows sharp peaks 
at two frequencies: $\omega_2=$ 163.7 \kmskpc\ and $\omega_1=$ 199.4 \kmskpc. 
These peaks contain about 95\% of total spectral power. Most of the remaining 
power is in secondary peaks at frequencies which are functions of $\omega_1$ 
and $\omega_2$, namely $\omega_1 + \omega_2$, $2\omega_1$, $2\omega_2$, 
$\omega_2-\omega_1$ etc. The power in the secondary peaks rapidly decreases
as a function of their frequency: the group of three peaks between 300 and 400 
\kmskpc\ contains less than 2\% of the total power, and further peaks contain
still one order of magnitude less of the total power.

The two frequencies, $\omega_1$ and $\omega_2$, at which almost all of the 
power resides, are separated by 35.75 \kmskpc. This is exactly twice 
$\Omega_2 - \Omega_1$, the difference between the pattern speeds of the bars 
in Model 1. In Sect.3, we showed that in the linear approximation 
the solution for the radial coordinate (\ref{r1f}) of a particle moving in the 
potential of two independently rotating bars, in the absence of free
oscillations, consists of two oscillations with frequencies $\omega_1$ and 
$\omega_2$ defined by (\ref{omom}). The difference between those frequencies 
is also $2(\Omega_2 - \Omega_1)$. Thus, we imply that the frequencies of the 
highest peaks in the power spectrum of the trajectory from Fig.2 are in fact 
the frequencies of driving by the two bars, and that they are nonlinear 
equivalents of the $\omega_i$'s defined by (\ref{omom}). From (\ref{omom}) 
one can  obtain the underlying orbital frequency of the particle: 
$\Omega =$ 123.75 \kmskpc. It is close to its epicyclic approximation at
the guiding radius $r_0 = (\Sigma r^i) / N = 1.803$ kpc, which is 
121.5 \kmskpc. This consistency further confirms our finding. Thus in both the 
linear case (epicyclic approximation) and in the general case, we constructed 
double-frequency orbits in an oscillating potential of a double bar, with 
frequencies equal to the frequencies of driving by the bars. Note that since 
the Fourier analysis requires the particle to be followed for many relative
periods of the bars, and since no fine adjustments of initial conditions were
made, this double-frequency orbit is most likely stable.

\subsubsection{Trajectory trapped around double-frequency orbit}
Below, we will show that the trajectory from the right panels of Fig.1 
involves free oscillation in addition to the two oscillations driven by
the bars. The power spectrum for this trajectory is shown in the middle row
of Fig.4. It shows four major peaks. The two highest peaks at $\omega_2=$ 
195.9 \kmskpc\ and $\omega_1=$ 231.7 \kmskpc\ are separated by twice the 
difference between the pattern speeds of the bars, $2(\Omega_2 - \Omega_1)$, 
which indicates that they occur at the frequencies of oscillations driven by
the bars. Again, from (\ref{omom}) one can obtain the underlying orbital 
frequency of the particle: $\Omega =$ 139.8 \kmskpc, and from averaging the 
radial coordinate one can get an approximation for the guiding radius
$r_0 = 1.579$ kpc. In the linear approximation, the orbital frequency 
at this guiding radius is 137.5 \kmskpc, again consistent with the nonlinear
solution. About 70\% of the total power resides in these two peaks at 
frequencies of the driving by the two bars.

Between the two highest peaks in the power spectrum, there are two other peaks 
of lower amplitude. They are located at frequencies $\kappa = $ 211.9 \kmskpc\ 
and $\kappa' =$ 215.5 \kmskpc, which are not commensurate with $\omega_1$ and 
$\omega_2$, and they contain about 20\% of the total power. Note that 
$\omega_1 - \kappa' = \kappa - \omega_2$, which means that $\kappa$ and 
$\kappa'$ are not independent, and therefore that these two peaks indicate 
only one additional frequency on the trajectory. We postulate that this
frequency is the free oscillation frequency. In the linear 
approximation of Sect.3, this frequency corresponds to $\kappa_0$ in 
(\ref{r1f}), which for the guiding radius $r_0 = 1.579$ kpc is equal to
202 \kms. This value is somehow different from $\kappa$ and $\kappa'$ 
measured in the power spectrum, but it remains in their vicinity.
Moreover, qualitatively one should expect that at this guiding radius $\kappa$
and $\kappa'$ lie in between $\omega_1$ and $\omega_2$, because in the linear 
approximation $\Omega - \kappa/2$ takes values between $\Omega_1$ and 
$\Omega_2$ there.

All other peaks in the spectrum occur at frequencies which are functions of 
$\omega_1$, $\omega_2$ and $\kappa$, and power contained in those peaks 
decreases rapidly with frequency. Thus it appears that the orbit from the 
right panels of Fig.1 is trapped around a double-frequency orbit 
and oscillates around it with the characteristic frequency $\kappa$.

\subsubsection{Chaotic trajectory}
Finally, the power spectrum of the trajectory from the left panels of Fig.1 
is shown in the bottom rows of Fig.4. There are more peaks, and also much power
contained in between the peaks. In order to validate the significance of the
peaks in this spectrum, we calculated again power spectra for the three 
trajectories from Figs. 1 and 2, but now followed for 600 alignments of
the bars, with dynamical variables recorded at $2^{16}$ equal time intervals. 
While the power spectra for the two trajectories analyzed above do not show
significant changes, the new spectrum for the trajectory from the left panels 
of Fig.1, marked with the thick line in the bottom-left panel of Fig.4, shows 
multitude of peaks, each of them containing no more than 2\% of the total 
power. Kandrup, Eckstein \& Bradley (1997) showed that a multitude of 
frequencies indicates chaos (see also Athanassoula 2005). In the case 
analyzed here, there is a continuous distribution of power with broad maximum 
in the frequency range between 120 and 190 \kmskpc. Such a power spectrum is 
characteristic of a chaotic trajectory. 

\subsection{The vicinity of a double-frequency orbit}
In order to study how particles get trapped around double-frequency orbits,
one can explore a range of initial conditions around the ones that characterize
it. We developed an automatic procedure that finds local maxima (peaks) in the 
power spectrum at frequencies $\omega_1$ and $\omega_2$ corresponding to the 
driving by the two bars and at the free oscillation frequency. We based this 
procedure on the fact that the two frequencies of driving by the two bars are 
always separated by $2(\Omega_2 - \Omega_1)$, where $\Omega_1$ and $\Omega_2$ 
are the (constant) pattern speeds of the two bars. Thus, the algorithm first 
searches for the highest peak in the power spectrum, and records its frequency 
$\omega$. Then it examines power spectrum at frequencies 
$\omega \pm 2(\Omega_2 - \Omega_1)$. If there is a peak in one of these 
frequencies, then this frequency and $\omega$ are recorded as
$\omega_1$ and $\omega_2$ on this orbit, as defined in (\ref{omom}). 
Furthermore, the free oscillation frequency $\kappa$ is found by searching 
for the peak with highest power that does not occur at a frequency which is 
a linear combination of $\omega_1$ and $\omega_2$.

\begin{figure}
\centering
\vspace{-2mm}
\includegraphics[width=1.05\linewidth]{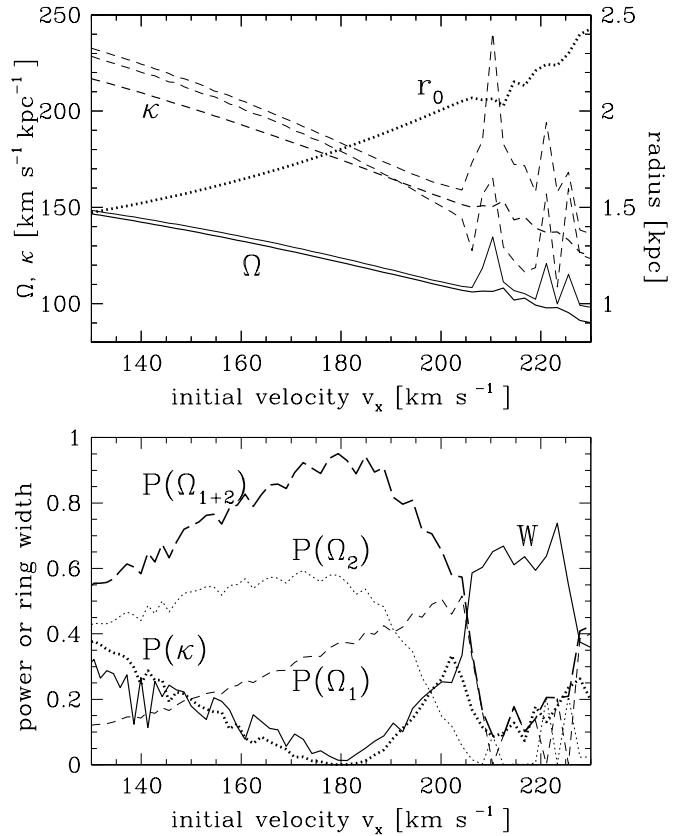}
\vspace{-6mm}
\caption[]{Parameters that characterize orbits in Model 1 by Maciejewski \& 
Sparke (2000), which start at 1.7-kpc distance from the galaxy's centre, on 
the minor axis of the bars when the bars are aligned, and with a
velocity vector at right angle to it, plotted as a function of the
magnitude of the starting velocity $v_x$.  
{\it Upper panel:} Thick dotted line marks the 
average radius of the orbit, $r_0 = (\Sigma r^i) / N$. The epicyclic 
approximation for the angular frequency $\Omega$ and for the frequency of free 
oscillations $\kappa$ at this radius are plotted with thick solid and dashed
lines, respectively. The values of $\Omega$ and $\kappa$ derived from the
power spectrum of the orbit are plotted with thin solid and dashed lines, 
respectively. {\it Lower panel:} Percentage of the total power residing at 
the frequencies of the major peaks. The thin dashed line marks the power 
$P(\Omega_1)$ at the driving frequency of the outer bar, and the thin dotted 
line marks $P(\Omega_2)$ at the driving frequency of the inner bar. The sum
of $P(\Omega_1)$ and $P(\Omega_2)$, $P(\Omega_{1+2})$, is plotted with a thick
dashed line. The thick dotted line marks the power $P(\kappa)$ in the two
peaks arising from the free oscillation frequency. The thick solid line marks
the median width of the ring enclosing particle positions at consecutive 
alignments of the bars, in units of the average radius of the orbit $r_0$ (see
Sect.4.4).}
\label{f5}
\end{figure}

In our analysis, we kept focus on a particle that starts at the 1.7-kpc 
distance from the galaxy centre, on the minor axis of the aligned bars, 
with its initial velocity perpendicular to that axis. We analyzed a range 
of values of the initial velocity between 130 \kms\ and 230 \kms. The initial 
conditions for the three trajectories studied above fall within this range.
The dotted line in the upper panel of Fig.5 indicates the average radius 
$r_0 = (\Sigma r^i) / N$ (where the meaning of the symbols is the same as 
in Sect.4.2) as a function of the initial velocity $v_x$. As 
expected, higher initial velocity results in larger average radius, although
the relation becomes not monotonic and noisy at initial velocities above
205 \kms. When the average radius is taken as the
radius of the guiding centre in the linear approximation, one can calculate
the orbital frequency $\Omega$ and the free oscillation frequency $\kappa$
in this approximation. They are drawn in thick solid and dashed lines, 
respectively. 

On the other hand, once we derive $\omega_1$ and $\omega_2$ from the power 
spectrum of the trajectory, the orbital frequency $\Omega$ in the full 
nonlinear solution can be obtained from (\ref{omom}). The values of $\Omega$
derived from the measured values of $\omega_1$ and $\omega_2$ are the same,
and they are plotted with a thin solid line in the upper panel of Fig.5.
They closely follow the values derived in the linear approximation for most of
the range, although there are clear departures at initial velocities above 
205 \kms. Although our algorithm to analyse the power spectrum allows us to 
extract automatically the free oscillation frequency $\kappa$, one should 
also expect a peak at $\kappa' = \omega_1 + \omega_2 - \kappa$, because two 
frequencies, $\omega_1$ and $\omega_2$, are already present in the system. In 
fact, the algorithm finds peaks at two frequencies $\kappa$ and $\kappa'$, 
whose sum is $\omega_1 + \omega_2$. Both of these frequencies are drawn in the 
upper panel of Fig.5 (thin dashed lines), since from the power spectrum one 
cannot determine which one is the actual $\kappa$. Their functional dependence
on the initial velocity $v_x$ is similar to that obtained in the linear
approximation, although the slope is slightly different, and the quantitative
discrepancy is larger than in the case of the orbital frequency $\Omega$.
Again, $\kappa$ and $\kappa'$ as a function of the initial velocity become
ragged for $v_x>$ 205 \kms. 

One can also examine how much of the total power resides in various peaks 
in the Fourier spectrum. In the lower panel of Fig.5, a thin dashed line
indicates the power $P(\Omega_1)$ residing in the peak at the frequency of 
driving by the outer bar, while a thin dotted line indicates the power 
$P(\Omega_2)$ in the peak at the frequency of driving by the inner bar. With
the thick dashed line we plotted the power $P(\Omega_{1+2})$ residing in both
peaks above. Power $P(\kappa)$, residing in the two peaks related to the free 
oscillation 
frequency, is plotted with a thick dotted line. From the plot one can see that
in the range of initial velocity $v_x$ between 130 \kms\ and 205 \kms,
$P(\Omega_{1+2})$ and $P(\kappa)$ sum up roughly to 100\%. This means that 
only three frequencies are present on those trajectories, as expected for the 
regular orbits (see Sect.2). In this range, the larger the value of 
$P(\kappa)$, the smaller $P(\Omega_{1+2})$. However, throughout this range
$P(\Omega_{1+2})> 0.5$, which means that more than half of the total power
is located in the two peaks at the driving frequencies of the two bars. Orbits 
with larger average radii $r_0$ have more power in the driving frequency of
the outer bar, while orbits with smaller $r_0$ have most of their power in the 
driving frequency of the inner bar. For $v_x$ around 180 \kms, up to 95\% of
the total power resides at the driving frequencies of the two bars, and only
0.03\% in the free oscillations. 
Thus the trajectory with $v_x$ around 180 \kms\ is made up almost entirely
of oscillations with the driving frequencies of the two bars. This is the
double-frequency orbit from Fig.2. The more $v_x$ differs from 180 \kms, the
larger fraction of the total power resides in the free oscillations: up to 
38\% and 33\% at $v_x=130$ \kms\ and $v_x=202$ \kms, respectively. This is 
typical for trajectories trapped around stable regular orbits. An example of 
such a trajectory is the one 
from the right panels of Fig.1, for which $v_x=150$ \kms.

For $v_x>$ 205 \kms, the summed power in peaks related to the three frequencies
considered here is far smaller than 100\%. The initial conditions of 
the trajectory from the left panels of Fig.1 belong to this range of $v_x$. The
power spectrum of that trajectory (Fig.4, bottom row) shows the power 
distributed in a continuum of frequencies, which is characteristic for a 
chaotic orbit. Thus trajectories with $v_x>$ 205 \kms\ are most likely chaotic. 
This explains the noise for $v_x>$ 205 \kms\ in the curves that mark the 
frequencies derived from Fourier analysis in the top panel of Fig.5. One can 
point out the region of onset of chaos at $v_x$ between 200 \kms\ and 210 
\kms, where both $P(\Omega_{1+2})$ and $P(\kappa)$ drop sharply with $v_x$.

\subsection{Construction of loops from double-frequency orbits}
In the epicyclic approximation, the double-frequency orbits have a nice
feature, namely that particles following them populate loops: closed curves 
that return to their original shape and position at every alignment of the 
bars. One may therefore expect that also in the general case considered here
particles on double-frequency orbits populate loops. In order to check, whether
it is the case, we follow the same procedure as in the epicyclic approximation.
We start a particle from the minor axis of the aligned bars, and record its
position and velocity every time the bars align again. Let {\bf $x_0$} mark the
starting position and velocity of the particle, {\bf $x_1$} its position and 
velocity at the first consecutive alignment of the bars and so on. Positions 
of a particle on a double-frequency orbit are overplotted in Fig.2 for the 
first 20 alignments of the bars. In the left panel of Fig.6, we plotted 60 
positions for 60 alignments. They seem to be arranged on a closed curve of 
oval shape.

\begin{figure}
\centering
\vspace{-2.3cm}
\includegraphics[width=0.99\linewidth]{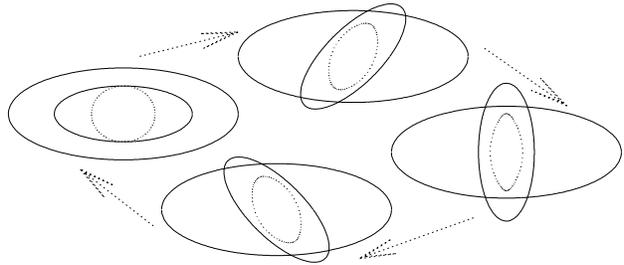}
\vspace{-2.5cm}
\caption[]{{\it Left panel:} Positions of a particle on the double-frequency 
orbit from Fig.2 at consecutive alignments of the bars generate a loop. Here
60 positions are recorded. {\it Remaining panels:} Evolution of the loop during 
one relative period of the bars. The bars, outlined with solid lines, rotate 
counterclockwise.}
\label{f6}
\end{figure}

Let us now take $N$ values of $x_i$ as initial conditions for a set of $N$ 
particles at one alignment of the bars. By construction, a particle starting 
at {\bf $x_i$} will be located at {\bf $x_{i+1}$} at the next alignment 
of the bars. Thus, if positions of a particle at every alignment of the bars
populate a closed curve, this curve will transform into itself every time
the bars align. This is what we defined as a loop. In the left panel of Fig.6
one can see that positions of a particle on a double-frequency orbit, written 
at every alignment of the bars, populate a closed curve, which is a loop. 
Therefore, one may expect that the relation between double-frequency orbits
and loops found in the linear approximation also holds in the general case.
Note that this construction of a loop requires the particle to be
followed for many relative periods of the bars, hence it will likely produce
loops that are maps of stable double-frequency orbits only.

The loop returns to its original shape and position at every alignment of the 
bars. Recording positions of particles from the loop at any time between
alignments of the bars will unveil how the loop evolves between the alignments.
This has been shown in the remaining panels of Fig.6, where we plotted 
locations of these particles at three other relative positions of the bars. 
The particles remain confined to the loop, but the shape of the loop varies 
in between the alignments.

Will particles on other orbits, which involve free oscillations, take any
particular positions at every alignment of the bars? One may expect that
since such orbits oscillate around double-frequency orbits, the recorded
positions will gather around the loop. If so, they should be enclosed within 
a band of certain width that increases with the amplitude of free oscillation.
This band, surrounding the loop, is closed and, following the notation
introduced by Maciejewski \& Sparke (2000), we will call it the ring.
In order to see how the width of the ring changes with the amplitude of free
oscillation, we applied the following algorithm, similar to the one used by 
Maciejewski \& Sparke (2000).
We registered particle positions on the orbit at 400 alignments of the
bars. We divided the full $2\pi$ angle into 80 equal parts, and for each part
we calculated the range $\Delta r$ of radii among the positions
falling into this part. The median of the $\Delta r$ values is taken
as the estimate of the ring width $W$. 

In the bottom panel of Fig.5 we plot the ring width $W$, normalized to the 
average radius of the orbit $r_0$, as a function of the starting velocity. At 
velocities between 130 \kms\ and 200 \kms, the ring width $W$ is below 30\%
of the average radius, hence recorded positions occupy a well defined ring.
The ring gets thinner for velocities close to 180 \kms. This is the initial
condition for the double-frequency orbit from Fig.2 that generates the loop
from Fig.6. For this orbit $W$ is about 2\%. Generally, in the range of 
initial velocities considered here, $W$ 
behaves in a very similar way to $P(\kappa)$. The larger the free oscillations
around the double-frequency orbit, the wider the ring enclosing recorded
positions. Therefore, we confirm our expectations that orbits trapped around
double-frequency orbits map onto rings surrounding the loop.

Note that in this algorithm, $W$ is zero only for circular loops or under 
infinitely fine sampling in angle. Otherwise, $W$ is only near zero for a 
loop. However, unlike the amplitude of free oscillation with frequency
$\kappa$, $W$ is very
straightforward to measure and to visualize. The excellent agreement between 
the functional behaviour of $W$ and $P(\kappa)$ exemplified in Fig.5 makes $W$ 
a useful indicator of a loop.

For initial velocities between 200 \kms\ and 210 \kms\ the median ring width 
sharply rises from 30\% to 70\% of the average radius. Such large median width 
indicates that recorded positions are no longer confined by a ring, but rather
densely populate some fraction of the plane, and do not gather 
on any curve. This is characteristic for chaotic orbits, which is consistent 
with the frequency analysis above.

\section{Discussion and conclusions}
The goal of this paper is to present an argument that the backbone of an 
oscillating potential of a doubly barred galaxy is built out of 
double-frequency 
orbits. In Sect.3, by using the linear approximation, we showed 
that two frequencies are indispensable in motion in such systems. A single
frequency, and thus a closed periodic orbit, can occur only when these two
frequencies are commensurate. However, even in such case they are two
separate frequencies, which happen to have the same numerical value. This is
a condition for resonance, like the one in the single bar when the frequency
of free oscillations is equal to that of driving by the bar.

Double-frequency orbits in the oscillating potential of double bars do 
not close. Their appearance is also different in the reference frame of
each bar. However, their perception and analysis is facilitated greatly by
the fact that they map onto loops, which are closed curves. The mapping is
done by recording the positions on the orbit at every alignment of the bars.
In such a construction, points that are on the loop will remain on the loop.
A loop can be viewed as a set of particles, and therefore its appearance is
independent from the frame of reference. However, the loop oscillates in time
with a period equal to the relative period of the bars.

In the linear approximation we showed rigorously that each double-frequency 
orbit maps onto a loop. In the general, non-linear case, we found that when 
the Fourier analysis
of the orbit indicates almost all power at the frequencies of driving by the
bars, and almost no power at the the free oscillation frequency, the map of
the orbit is to a good approximation a 1-D curve. This is the loop. When free 
oscillations are present, the orbit maps onto a set of points in a ring 
surrounding the loop. The width of this ring is higher when more power is in 
the free oscillations.

From the power spectrum of an orbit in the oscillating potential of a doubly barred
galaxy we can retrieve its three fundamental frequencies (frequencies of 
forcing by the two bars and the frequency of free oscillations) when the orbit
is regular. Otherwise a continuous power spectrum is recovered when the orbit 
is chaotic. The Fourier algorithm that we apply displays the baseline at no 
more than $10^{-4}$ of the total power, hence peaks higher than that can
be easily extracted. Thus, even if the power in e.g. the free oscillation 
is very small around the loop, our algorithm recovers its frequency correctly, 
as the continuity of the $\kappa(v_x)$ curve in the top panel of Fig.5 
indicates. In our algorithm, the power spectrum of each trajectory is analyzed 
independently from power spectra of neighbouring trajectories, and frequencies 
recovered on one trajectory do not enter as an `initial guess' for those on 
the nearby trajectory. Therefore, continuity of $\Omega$ and $\kappa$ as 
functions of the initial velocity $v_x$ is a genuine property of the system, 
and not of the algorithm.

Methods employed in this paper in the search for double-frequency orbits and
loops require the particle to be followed for many (i.e. hundreds) relative
periods of the bars. We do not perform fine tuning of the initial conditions 
on the trajectory, hence recovering unstable orbits with these methods is 
rather unlikely. Therefore double-frequency orbits recovered with the methods 
employed in this paper are stable in the sense that they trap regular orbits,
which wind tightly around them and map onto rings enclosing loops.

We conclude that
in a potential of two independently rotating bars, regular trajectories are 
trapped around parent stable double-frequency orbits. They oscillate around 
these orbits, like trajectories in a single bar oscillate around stable 
closed periodic orbits. The two frequencies present on the parent 
orbits correspond to the frequencies of driving by the two bars. Therefore 
the parent orbits do not close in any reference frame. However, they can 
be mapped onto the loops, whose appearance is independent from the reference
frame. As such, loops are a unique tool that greatly facilitates the search for
regular orbits in oscillating potentials.

\vspace{5mm}

{\bf Acknowledgments.} This work was carried out within the framework of the 
European Associated Laboratory ``Astrophysics Poland-France". It was 
supported by the Polish Committee for Scientific Research as a research 
project 1 P03D 007 26 in the years 2004--2007. WM thanks the CNRS for
an associate position, which made this collaboration possible.

\end{document}